\documentclass[a4paper,11pt]{article}
\usepackage{jcappub} % for details on the use of the package, please see the JINST-author-manual
\begin{document}
%\arxivnumber{1234.56789} % Only if you have one %%UBACI KADA DOIJES BROJ ZA SUBMISIJU U CASOPIS
\title{\boldmath Bounding the graviton mass using non-linear density wave theory}

\author[1]{M. Vukcevic\note{Corresponding author.}}

\affiliation[1]{Astronomical Observatory Belgrade,\\Volgina 7, Serbia}

\emailAdd{vuk.mira@gmail.com}

\abstract{In this paper we use the Newtonian gravitational potential corrected by non-liner effects to obtain new bounds on graviton mass using non-linear density wave theory (NLDW).
This potential differs from the gravitational potential obtained in other modified gravity theories (e.g. the weak field limit of Yukawa gravity, Modified Newtonian Dynamics, non-local theories, $\Lambda$ cold dark matter..). 
Using this model, we are able to define wavelength of the non-linear wave as an analytical solution of integrable non-linear differential equation (namely, non-linear Schrodinger equation). Assuming that the wavelength of the non-linear wave represents the graviton Compton wavelength, we have found the corresponding upper bound of graviton mass. We compare obtained result with first assessments of  LIGO $\&$ Virgo collaboration and we find they are in a good agreement.
Present model used to determine the upper limit of graviton mass is completely independent from other methods published until now. We have compared our result with results obtained using several chosen published methods.}

\maketitle

\flushbottom

\section{Introduction}
\label{sec:intro}

Since 2016 the LIGO \& Virgo collaboration presented the first discovery of gravitational waves (GW) in merging black holes, altogether with an upper limit on graviton mass such as $m_{g}< 1.2 \times 10^{-22}$ eV \citep{aabb16a}, the search for an adequate theoretical model has been amplified. 
Although their observational data showed that there is no any violation of classical general relativity (GR), 
several perspectives occurred for cosmology  \citep{aabb17a}, sources of GW  \citep{aabb18} and testing the the standard theory of gravity in regimes being out of scope previously.

Recent simultaneous measurements of arrival times in strongly lensed GW signals and associated electromagnetic counterparts give an opportunity to estimate GW propagation speed  \citep{aabb16b},  \citep{bran16}. These time delays depend on the matter distribution in the lens (galaxy), on the overall matter distribution along the line of sight and on the cosmological parameters  \citep{coll17},  \citep{fan17}.

On the other side, it was recently proposed that dark matter interacting with GW can produce shear viscosity which can be used as a mechanism to explain the accelerated expansion of the Universe.
However, neither of proposed models, including standard model of elementary particles do not provide any satisfactory explanation for two huge discrepancies between observations and expectations based on the theoretical model. One of many tensions is inability of CDM model to fully explain dark matter and dark energy phenomena as a complete theory. 
A number of alternative models have been proposed in literature, like those based on a modification of the fundamental laws of gravity: MOND  \citep{mof05},  \citep{mil20}, non-local theories  \citep{dim20} or Yukawa gravity potential  \citep{wil98},  \citep{san86}. The weakness of all these theories is pure mathematical form with no physical background. Also, neither of these theories is unique on wide spatial scale range.

Other frameworks postulates non-zero graviton rest mass  \citep{gold10},  \citep{rham17} within which the mass of this particle has been incorporated into the appropriate gravity theory. 
When the graviton is given a mass, it is necessary to modify Newtonian potential. However, any of proposed theories faces
sever problem related to flat rotation curves of stars in galaxies  \citep{rub20},  \citep{ber18}, as well as with the stability of galaxy clusters  \citep{zwik33},  \citep{smi36},  \citep{hag13} suggesting the existence of much more matter than visible one. Without entering details on the dark matter problems, we propose here new approach based on the non-linear galaxy dynamics used to determine graviton mass. This model is extension on the gravity potential, based on Newtonian gravity, which is related to the natural occasion of spiral structure formed due to dispersion balanced by non-linear effects. This balance is consequence of the marginal stability of the galactic disk, resulting in the soliton wave existence. 

The content of this paper is as follows. In Section 2, we present a model used to estimate Compton wavelength. In Section 3, we estimate graviton mass based on non-linear gravity potential, explaining details of this approach.
Last section is devoted to comparison with other models and discussion.

%%%%%%%%%%%%%%%%%%%%%%%%%%%%%%%%%%%%%%%%%%%%%ovo ide posle za komparison

%%%%%%%%%%%%%%sad ovde model za compton wavelength%%%%%%%%%%%%
\section{Method} \label{sec:meth}

A galaxy is a N-body system, composed of stars and gas, interacting with each other through a long range gravity force. There are a lot of common characteristics between the physics of galaxies and plasmas, systems of charged particles interacting through the Lorentz force. Macroscopic description of these systems can be well described by a fluid model with incorporated Poisson's equation which includes collective behaviour, such as waves or instabilities.

Linear density wave theory done by Lin and Shu  \citep{lin64} explained well spiral pattern formed within a disk.  Within the infinitesimally thin disk approximation, it has been obtained a spiral wave solution, but this theory failed to explain long lasting structure due to differential rotation that would wind up the pattern in time scale insufficient to be observed. There were  several attempts to overcome this problem  \citep{toom69},  \citep{korm79},  \citep{kend11},  \citep{purc11}. However, our approach extend Lin \& Shu model keeping nonlinear terms in expanded variables. 
The density wave model is based on the fluid description of the galactic disk. Namely,  there are transport
equations for the mass density $\rho$ and the momentum
$\rho v$, together with the Poisson's equation that relates
the density to the gravitational potential $\phi$. We keep  same notation as in work done in  \citep{vuk14}
%\cite{vuk14}
for the stellar disk. 

The equilibrium state of the system is described as a rotation
with an angular velocity $\Omega(r)$ about z-axis under
the balance of centrifugal, gravitational and pressure forces in
a frame rotating with constant angular velocity $\Omega_{0}$.
Then, the equilibrium velocity is $v_{0\varphi} = (\Omega - \Omega_{0})r$,
where $\Omega^{2} r = -\partial\phi_{0}/\partial r$. All quantities with subscript $0$ are reserved for the equilibrium functions. Pressure has been neglected for the simplicity reason since the gas follow the same pattern slightly inclined to the stellar one  \citep{vuk24}.

The dispersive property originates from the
coupled Poisson's equation, which is a second-order
elliptic partial differential equation. At the same time, this particular equation defines the geometry of the system. In the infinitesimally thin disk approximation, Lin and Shu assumed
delta function for the density in z-direction and approximates
Poisson's equation by

\begin{equation}
	\frac{\partial\phi (r,z=0)}{\partial r}=\pm 2\pi iG\sigma,	
\end{equation}
where $\sigma$ is surface
mass density  \citep{lin64}.
%\cite{lin64}. 
Than, relation between
surface density and two-dimensional potential is 

\begin{equation}
\sigma=-\frac{k}{2\pi G}\phi(z=0)	,	
\end{equation}
where  wave number is $k=-\frac{i}{\phi}\frac{\partial\phi}{\partial r}$  \citep{lin64}.

In order to derive possible non-linear equation, it is necessary to discuss parameter regime defined by dispersion equation. It will  determine
the unique transformation of coordinates and expansion of variables within the Reductive Perturbation Method (RPM)  \citep{jaf64}. Here, we are not going to enter in details of the procedure since it has been already done, we rather list implications of such calculation.

For differentially rotating thin stellar disk, linearised  fluid equations altogether with Poisson equation, assuming plane wave type variation as $f=f(r)e^{i(kr+m\varphi - \omega t)}$, leads to the dispersion relation 

\begin{equation}
	(\omega-m\Omega)^{2}=\kappa^{2}-2\pi G\rho_{0}|k|,
\end{equation}
where $\omega-m\Omega$ is Doppler shifted frequency,  $m$ is azimuthal wave number, and $\kappa$ is epicyclic frequency due to
differential rotation

\begin{equation}
	\kappa^{2}=2\Omega(2\Omega +r(\frac{d\Omega}{dr})).
\end{equation}

\subsection{Marginal stability} 

Using eq. 3, stability parameter is defined by $k_{2}=\frac{\kappa^{2}}{2\pi G\rho_{0}}$, where $k_{2}$ is normalised by $\frac{\kappa^{2}}{2\pi G\rho_{0}}$, so all waves with $k<k_{2}$ are purely stable. 
Initial limitation on the wave number $k>k_{1}$ regarding vertical stability of the disk, with $k_{1}=max \left\{\frac{1}{r},\frac{\rho_{0}'(r)}{\rho_{0}(r)}\right\}$ where sign ' denotes derivative with respect to r, provides lower limit of the wavenumber; observational data suggests that $k_{1}\approx k_{2}$ in real galaxy  \citep{ber00}
%\cite{ber00} 
defining the marginal stability of the galactic disk.
In this marginal stability case, special transformation of variables has to be introduced, different from the stable case due to the condition that 
frequency goes to zero, and consequently group velocity becomes infinite  \citep{wat69}. 

Using this transformation of coordinates
with a perturbation expansion of the dependent variables, one can obtain single nonlinear
equation (NLS). This type of perturbation has been developed and formulated in
general way by Taniuti and his collaborators  \citep{jaf64}. %\cite{jaf64}.

In the case of galaxy, it has been done by   \citep{vuk14},  \citep{vuk24} and we skip here derivation details. For the purpose to estimate graviton mass, it is relevant marginal stability condition in terms that wavenumber $k\sim1$. So that, 

\begin{equation}
	k=\frac{2\pi G \rho_{0}}{\kappa^{2}} \frac{2\pi}{\lambda_{g}}\approx1,
	\end{equation}
leads to the value of characteristic length scale (Compton  wavelength) $\lambda_{g}=10^{17}$ cm after substituting typical values for Milky Way $\rho_{0}=4\times10^{-2}g/cm^{2}$ and $\kappa=10^{-15}1/s$  \citep{vuk21}.

%%%%%%%%%%%%%%%%%%%%%%%%%%%%%%%%%%%%%%%%%%%%%%%%%
\section{Graviton mass} \label{sec:gravmas}

Graviton is considered to be the carrier of the gravitational interaction ,so that spin-2 tensor boson, electrically uncharged. Regarding mass, it is massless since, according to GR, it travels along null geodesics like photon, at the speed of light c. 
According to some alternative theories, gravity is propagated by a massive field, by a graviton with some small, nonzero mass $m_{g}$ (first introduced by Fierz and Pauli  \citep{fie39}).  However, theories of massive gravity were able to possible explain the accelerated expansion of the Universe without dark energy (DE) hypothesis, and important predictions that the velocity of gravitational waves should depend on their frequency. So that, the effective gravitational potential should include a correction depending on the Compton wavelength of graviton $\lambda_{g}= h/(m_{g} c)$. In the other words, if gravitation is propagated by a massive field, then the effective Newtonian potential has, for example, a Yukawa form: $\sim r^{-1}e^{(r/\lambda_{g})}.$ Then, the massive graviton propagates at an energy E dependent speed $v^{2}_{g}/c^{2}=1-m^{2}_{g}c^{4}/E^{2}=1-h^{2}c^{2}/\lambda^{2}_{g}E^{2}$ or, equivalently frequency f, $v^{2}_{g}/c^{2}=1-c^{2}/f\lambda^{2}_{g}$  \citep{wil98},  \citep{wil14}. 

The above modified dispersion relation is obtained from the modified special relativistic relation between energy E and momentum p of graviton $E^{2}=p^{2}c^{2}+m^{2}_{g}c^{4}$, with its velocity $v_{g}$ satisfying $v_{g}/c=cp/E$. Using time difference between GW and electromagnetic wave emitted from the same object $\bigtriangleup t=\bigtriangleup t_{a}-(1+z)\bigtriangleup t_{e}$, where $\bigtriangleup t_{a}$  and $\bigtriangleup t_{e}$ are differences in arrival and emission time of these two signals, and $z$ is the redshift of the object, one can obtain the observational constraint on the $v_{g}/c$ ratio. Using $\bigtriangleup t$, sped difference reads as $1-v_{g}/c=5\times10^{-17}(200Mpc/D)(\bigtriangleup t/1 s)$, where $D$ is distance of the object  \citep{wil98}.

Here, we present example of Yukawa gravity theory and its application on the constraint of Compton wavelength and graviton mass; it will be used as comparison with our model.
In the Yukawa-like potential of the form $U(r)=(G_{\infty}M/r)(1+\alpha e^{-r/r_{0}})$, the strength of interaction $\alpha$ is not unique value but rather define the scale of interaction, so that gravitational constant measured at infinity $G_{\infty}$ and localy $G_{0}$ are related via $\alpha$ as $G_{0}=G_{\infty}(1+\alpha)$, while $r_{0}$ is characteristic length scale. 

If the length scale $r_{0}$ corresponds to graviton mass $m_{0}$ as $r_{0}=h/m_{0}c$
then the flat rotation curves of spiral galaxies could be accounted for $\alpha\sim-1$
without introducing dark matter (DM) hypothesis  \citep{san86}. The negative sign of the strength of Yukawa interaction indicates an additional repulsive (anti-gravity) force which could mimic the effects of dark energy (DE) on the large scales. However, weakness of this theory is in fact that correction $\alpha$ has to change the value for short (Solar system, close binary system) and long scales (galaxy, galaxy cluster).

On contrary, our model accounts on the physical background of spiral galaxy dynamical condition, marginal stability, that is used to find out bound on graviton mass.

%%%%%%%%%%%%%%%%%%%%%%%%%%%%%%%%%%%%%%%%%%%%%%%
Using RPM in order to estimate nonlinear  effects in density wave theory done by Vukcevic  \citep{vuk14}, the gravity potential gradient $\partial \phi/\partial r$ is approximated by

\begin{equation}
	\frac{\partial \phi}{\partial r}=r\Omega^{2}+\sum^{n=1}_{\infty}\sum^{\infty}_{m=-\infty} 2\pi G \epsilon^{n}\Re(\rho^{(n,m)}(\xi,\eta)e^{i(kr-\omega\tau)}),
	\end{equation}
where term $r\Omega^{2}$ comes out from the equilibrium property. $\Omega$ is angular velocity, $G$ is gravitational constant, $\rho$ denotes surface mass density, $\xi$, $\eta$ and $\tau$ are corresponding stretched coordinates, while $k$ and $\omega$ are wave number and frequency, respectively.

In order to derive gravitational potential, we use already derived solution of Nonlinear Sr$\ddot{o}$dinger Equation for surface density perturbation:	
	\begin{equation}
	\rho^{1,1}(\xi,\eta)=\rho_{a}\frac{e^{i\psi}}{ch(\sqrt{\frac{Q}{P}}\rho_{a}(\xi-P\eta))}.
		\end{equation}
		
The wave phase $\psi$ is not relevant since only the real part of eq. 7 is taken, and parameters $P=\kappa/\pi G\rho_{0}=1/V_{g}$ and $Q=\kappa^{3}/\pi G\rho_{0}$ are related to the soliton velocity $V_{g}$ and width of the soliton; $\rho_{a}$ is wave amplitude and $\kappa$ is epicyclic frequency.

We substitute exact solution for density eq. 7 into equation for the potential gradient eq. 6, and after integration, the expression for non-linear gravitational potential reads as

\begin{equation}
	\phi(r)=\frac{\Omega^{2} r^{2}}{2}+\frac{ar}{ \cosh b(T-cr)}.
	\end{equation}
All parameters and variables are dimensionless. Returning to original coordinates, the non-dimensional velocity is multiplied by $2\pi G \rho_{0}/\kappa$, and $T=1=(t+\varphi/\Omega)$ by $\kappa$, where $\varphi$ is polar angle, $a=2\pi G \rho_{0}\rho_{a}[km/s^{2}]$, $b=\kappa\rho_{a}[1/s]$ and $c=1/V_{g}=\kappa / 2\pi g \rho_{0}[s/km]$. Evaluating time at $10^{9}$ yr (which is $3\times10^{16}$ s), taking $r$ in $[kpc]$ and group velocity $V_{g}$ in $[km s^{-1}]$, it is possible to derive the parameters $a$, $b$, and $c$ in eq. 8, which guarantee the flat shape of rotation curve even for radius order of $Mpc$ (see  \citep{vuk21},  \citep{vuk22},  \citep{vuk25}) without inclusion of DM.

Using eq. 5, we have derived $\lambda_{g}$ and from relation $\lambda_{g}=h/m_{g}c$ we can calculate $m_{g}=h/\lambda_{g}c$, where $h$ is Planck's constant. Therefore, $m_{g}=1.23\times10^{-21}$ eV, which is in good agreement with LIGO$\&$Virgo upper limit for graviton mass reported as $m_{g}<1.2\times10^{-22}$ eV.

\section{Conclusions}

Combination of gravitational wave domain and dynamical tests of the Yukawa correction of gravity potential was a very promising in testing the graviton mass.

LIGO\&Virgo Collaborations used the Yukawa correction to 
the gravitational potential with characteristic length scale $\lambda_{g}$ as $U(r)=(GM/r)(1-e^{-r/\lambda_{g}})$ 
in order to compare the so-called static and dynamical bounds on graviton mass  \citep{lig16}. 

The static bounds, such as those from the Solar System observations, do not probe the propagation of gravitational interaction, while dynamical bounds follow the very massive and compact objects, such as binary pulsars and supermassive Kerr black holes.  Dynamical bound reads $\lambda_{g}>10^{13}$ km at 90 \% confidence, which corresponds to a graviton mass $m_{g}<1.2\times10^{-22}$ eV, obtained from GW150914. It is for a factor of three better than the existing Solar-System static bounds  \citep{lig116}.
A new bound of $m_{g}<2.9\times10^{-21}$ eV has been derived using the simulations of the S2 star orbit at the Galactic Center  \citep{zak16}. Galaxy clusters give much stronger bounds $m_{g}<10^{-29}$ eV with corresponding $\lambda_{g}>10^{19}$ km  \citep{gup18}(Gupta and Desai, 2019). Estimates based on observations of dynamical properties of astrophysical objects at large scales provide one of the most stringent bounds on graviton mass compared to the measurements of the GWs speed  \citep{aabb16a},  \citep{aabb16b},   \citep{aabb21}, \citep{wil98}.

%%%%%%%%%%%%%%%%%%%%%%%%%%%%%%%%%%%%%%%%%%%%%%%%%
However, the approach considered in this work is different, independently on the scale, as far as details of modified gravity are concerned. As we have already mentioned above, this non-linear model is not modification, it is rather more accurate model of Newtonian gravity that accounts on the non-linear terms usually neglected as small, representing the multiplication of two perturbations. In some specific cases, when certain conditions are satisfied, these terms become comparable with dispersive ones and balancing each other they create so-called soliton wave (sometimes, it is used tsunami expression that we found inappropriate). Such a wave provides several advantages in galaxy dynamics research; here is used to estimate graviton mass as a consequence of observed pattern and dynamical parameters. Estimated value well agree with the value estimated by LIGO interferometer. 
Result based on nonlinear approach could be improved, or even better, it can be complemented by nonlinearity incorporated into larger scale dynamics, that will be subject of further research.

\section*{Acknowledgements}
Part of this research is supported by the Ministry of Education and Science of the Republic of Serbia (contract 451-03-66/2024-03/200002) and part from the project The Beijing Natural Science Foundation (NAR IS24021).

%%%%%%%%%%%%%%%%%%%%%%%%%%%%%%%%%%%%%%%%%%%%%%%%%%
%%%%%%%%%%%%%%%%%%%%%%%%%%%%%%%%%%%%%%%%%%%%%%%%%%%
%%%%%%%%%%%%%%%%%%%%%%%%%%%%%%%%%%%%%%%%%%%%%%%%%%

%\appendix
%\section{Some title}
%Please always give a title also for appendices.

%\acknowledgments

%This is the most common positions for acknowledgments. A macro is
%available to maintain the same layout and spelling of the heading.

%\paragraph{Note added.} This is also a good position for notes added
%after the paper has been written.


\end{document}